\documentclass{moriond}

\def\NIMA{{\em Nucl. Instrum. Methods} A }
\def\NPB{{\em Nucl. Phys.} B }
\def\PLB{{\em Phys. Lett.}  B }
\def\PRL{{\em Phys. Rev. Lett.}}
\def\PRD{{\em Phys. Rev.} D }

\def\HZZ{$\hbox{H}\rightarrow\hbox{ZZ }$}
\def\HWW{$\hbox{H}\rightarrow\hbox{WW }$}

\newcommand{\Photo}{}

\begin{document}
\vspace*{4cm}
\title{CMS High mass WW and ZZ Higgs search with the complete LHC Run1 statistics}

\author{M. Pelliccioni\\ for the CMS collaboration}

\address{Istituto Nazionale di Fisica Nucleare, Via P. Giuria 1, 10125 Torino, Italy}

\maketitle
\abstracts{
  A search for the decay of a heavy Higgs boson in the H$\to$ZZ and H$\to$WW channels is reported, analyzing several final states of the H$\to$ZZ and H$\to$WW decays. The search used proton-proton collision
  data corresponding to an integrated luminosity of up to 5.1 fb$^{-1}$ at $\sqrt{s} = 7$ TeV and up to 19.7 fb$^{-1}$ at $\sqrt{s} = 8$ TeV recorded with the CMS experiment at the CERN LHC. A Higgs boson with
  Standard Model-like coupling and decays in the mass range of 145 $< m_H <$ 1000 GeV is excluded at 95\% confidence level, based on the limit on the product of cross section and branching fraction.
  An interpretation of the results in the context of an electroweak singlet extension of the standard model is reported.
}

\section{Introduction}

In the Standard Model (SM) of electroweak (EW) interactions the existence of the Higgs boson, a scalar particle associated with the field responsible for spontaneous EW symmetry breaking~\cite{bib1,bib2} is predicted. The ATLAS and CMS experiments reported in 2012 the observation of a new boson with a mass of about 125 GeV \cite{bib3,bib4}. We refer to this newly observed Higgs boson as h(125) in this proceedings. While this boson shows SM-like properties, it is possible that it is merely part of a larger EW symmetry breaking sector. This can be accommodated in several extensions of the SM. In particular, we consider the scenario in which the SM Higgs boson mixes with a heavy EW singlet \cite{bib5}. This scenario is also useful to construct a general modelization of the Higgs sector that allows to interpret the data for several possible Higgs sector configurations.

Both ATLAS and CMS reported several searches for heavy SM-like Higgs bosons. In Ref.\cite{bib4}, ATLAS excludes a SM-like heavy Higgs boson in the mass range of 131 $< m_H <$ 559 GeV at 95\% CL. The CMS collaboration excluded an additional SM-like Higgs boson to to masses of 710 GeV at 95\% CL \cite{bib6}. None of the searches by ATLAS and CMS was performed using the full Run-1 LHC statistics collected by the collaborations.

We report on an extension of the CMS search using the full Run-1 dataset. In addition to the previous CMS analysis, we interpreted the data in the scenario of the SM expanded by an additional EW singlet. Both the possible SM-like heavy Higgs boson and the EW singlet are indicated as H. The analysis is performed using the proton-proton collision data recorded by CMS~\cite{bib24}, corresponding to integrated luminosities of up to 5.1 fb$^{-1}$ at $\sqrt{s} = 7$ TeV and up to 19.7 fb$^{-1}$ at $\sqrt{s} = 8$ TeV. The search is conducted in the 145 $< m_H <$ 1000 GeV mass range, exploiting both the \HZZ and the \HWW decay channels, which are the most sensitive to high mass Higgs boson decays. The lower boundary of the search is chosen to limit the contamination of h(125). In the \HZZ decays, we consider the final states containing four charged leptons (H$\rightarrow$ZZ$\rightarrow$2l2l'), two leptons and two neutrinos (H$\rightarrow$ZZ$\rightarrow$2l2$\nu$) and two leptons and two quarks (H$\rightarrow$ZZ$\rightarrow$2l2q), where l = e, $\mu$ and l' = e, $\mu$, and $\tau$. In the \HWW decays, we consider the fully leptonic (H$\rightarrow$WW$\rightarrow$l$\nu$l$\nu$) and semileptonic (H$\rightarrow$WW$\rightarrow$l$\nu$qq) decays.

\section{Simulation and Signal Modelization}

In order to simulate the signal and the background, we use several Monte Carlo event generators. For the Higgs boson signal, we generate samples for gluon-gluon fusion (ggF) and vector boson fusion (VBF) at next-to-leading order (NLO) using POWHEG 1.0~\cite{bib7,bib8} and a dedicated program for angular correlations~\cite{bib9}. Associated production of the Higgs boson with a vector boson (WH and ZH) and ttH are generated using PYTHIA 6.4 at leading order (LO)~\cite{bib10}. Events are weighted at generator level according to the total cross section of pp$\rightarrow$H~\cite{bib11}, which includes the ggF next-to-next-to-leading order (NNLO) and next-to-next-to-leading-log (NNLL) contributions, and the VBF NNLO contributions. The diboson invariant mass lineshape for signal is affected by the quantum interference between signal and the SM background. We correct the generated $m_H$ lineshape to obtain the theoretical predictions~\cite{bib12,bib13,bib14}.

The background from qq$\rightarrow$ WW production is generated with MADGRAPH 5.1~\cite{bib15}. The background from qq$\rightarrow$ZZ production is simulated with POWHEG at NNLO. The gluon gluon induced vector boson pair background (gg$\rightarrow$VV) is simulated at LO using GG2VV 3.1~\cite{bib16}. The other background processes considered (WZ, Z$\gamma$, W$\gamma$, W+jets and Z+jets) are generated using PYTHIA and MADGRAPH. Backgrounds from tt and tW events are generated with POWHEG at NLO.

PYTHIA is used for parton showering, hadronization and underlying event simulation for all the samples. The detector response is simulated using a detailed description of the CMS apparatus, based on the GEANT4 package~\cite{bib17}. Simulated samples include the presence of multiple proton-proton interactions per bunch crossing (pileup).

We test the presence of both a heavy SM-like Higgs boson and of an EW singlet scalar mixed with h(125). In the EW singlet scenario, the couplings of both states are constrained by unitarity and the coupling of h(125) is therefore lower than in the SM case. Unitarity is enforced by the relation $C^2 + C'^2 = 1$, where C and C' are the scale factors of the couplings of h(125) and the high mass Higgs boson, respectively, with respect to the SM. The production cross section modifier (also known as signal strength) and the width of the high mass Higgs boson are defined as

\begin{equation}
\mu' = C'^2 (1-\mathcal{B}_{new}),
\label{eqn1}
\end{equation}

\begin{equation}
\Gamma' = \Gamma_{SM}\frac{C'^2}{1-\mathcal{B}_{new}},
\label{eqn2}
\end{equation}

where $\mathcal{B}_{new}$ is the branching fraction of the EW singlet to non-SM decays. The signal strength measured for h(125)~\cite{bib18} can be used to put a 95\% CL limit on $C' < 0.28$.

We focus on the case where $C'^2 \leq (1-\mathcal{B}_{new})$, where the new boson will have a width equal or narrower with respect to the SM Higgs boson. We generate signal samples for different values of the width and scan the C' and $\mathcal{B}_{new}$ parameter space. In order to account for the proper signal and interference lineshape we follow the recommendations of the {\it LHC Higgs Cross Section Working Group}~\cite{bib11} (HXSWG). The interference between the high mass Higgs boson and the background is assumed to scale with the modified coupling of the Higgs boson. The interference between h(125) and the EW singlet partner is assumed to be small and is covered by a conservative systematic error~\cite{bib19,bib20}.

\section{Analyzed channels}

The results reported are obtained through the combination of different production and decay modes, as reported in Table~\ref{tab1}. All searches are restricted to the invariant mass region above 145 GeV, and for all the final states there are no events overlapping. For the \HWW$\rightarrow$ l$\nu$l$\nu$ decay, the EW singlet model interpretation starts at 200 GeV to avoid contamination from h(125). The \HWW$\rightarrow$ l$\nu$l$\nu$, \HWW$\rightarrow$ l$\nu$qq (with merged jets) and \HZZ$\to$ 2l2q decay channels are analyzed in the $\sqrt{s} = 8$ TeV sample only.
A detailed description of the analysis strategy for all the final states is provided in Khachatryan {\it et al.}~\cite{bib21}.

\begin{table}[t]
  \caption[]{Analyses included in this combination. The column ``H production'' indicates the production mechanism considered in the analysis. Untagged categories are mostly populated by ggF events. Events with a dijet pair consistent with a VBF topology are referred to as $(\mathrm{jj})_\mathrm{VBF}$. The category with dijet pairs and single merged jets from a Lorentz-boosted W (Z) are referred to as $(\mathrm{jj})_{W(Z)}$ and $(\mathrm{J})_{W(Z)}$ respectively. Three possible b-tag categories are identified with ``0,1,2 b tags''.}
\label{tab1}
\vspace{0.4cm}  

\begin{center}
\small
\begin{tabular}{cccccc}
\hline
 H & H & Exclusive & No. of & $m_{H}$ range & $m_{H}$ \\
 decay mode & production & final states & channels & [GeV] & resolution \\
 \hline 
$WW \to$ l$\nu$l$\nu$        & untagged  &  ((ee,$\mu\mu$),e$\mu$) + (0 or 1 jets) & 4  & 145--1000 & 20\%  \\
                             & VBF tag   &  ((ee, $\mu\mu$),e$\mu$) + $(\mathrm{jj})_\mathrm{VBF}$ & 2  & 145--1000 & 20\%  \\
\hline
$WW \to$ l$\nu$qq            & untagged  &  (e$\nu$, $\mu\nu$) + (jj)$_{W}$ & 2  & 180--600 & 5--15\%  \\
                             & untagged  &  (e$\nu$, $\mu\nu$) + (J)$_{W}$ + (0+1-jets) & 2  & 600--1000 & 5--15\%  \\
                             & VBF tag   &  (e$\nu$, $\mu\nu$) + (J)$_{W}$ + $(\mathrm{jj})_\mathrm{VBF}$ & 1  & 600--1000 & 5--15\%  \\
\hline
$ZZ \to$ 2l2l'               & untagged  &  4e, 4$\mu$, 2e2$\mu$ & 3  & 145--1000 & 1--2\%   \\
                             & VBF tag   &  (4e, 4$\mu$, 2e2$\mu$) + (jj)$_{\mathrm{VBF}}$ & 3 & 145--1000 & 1--2\%   \\
                             & untagged  &  (ee,$\mu\mu$) + ($\tau_h \tau_h$, $\tau_e \tau_h$, $\tau_\mu \tau_h$, $\tau_e \tau_\mu$) & 8 & 200--1000 & 10--15\%  \\
\hline
$ZZ \to$ 2l2$\nu$            & untagged  &  (ee,$\mu\mu$) + (0 or $\geq$ 1 jets) & 4 & 200--1000 & 7\%   \\
                             & VBF tag   &  (ee,$\mu\mu$) + $(\mathrm{jj})_\mathrm{VBF}$ & 2 & 200--1000 & 7\%   \\
\hline
$ZZ \to$ 2l2q                & untagged  &  (ee,$\mu\mu$) + $(\mathrm{jj})_{Z}^{0,1,2 \,\hbox{b tags}}$ & 6 & 230--1000 & 3\% \\
                             & untagged  &  (ee,$\mu\mu$) + $(\mathrm{J})_{Z}^{0,1,2 \,\hbox{b tags}}$ & 6 & 230--1000 & 3\% \\
                             & VBF tag   &  (ee,$\mu\mu$) + $(\mathrm{jj})_{Z}^{0,1,2 \,\hbox{b tags}}$ + $(\mathrm{jj})_\mathrm{VBF}$ & 6 & 230--1000 & 3\% \\
                             & VBF tag   &  (ee,$\mu\mu$) + $(\mathrm{J})_{Z}^{0,1,2 \,\hbox{b tags}}$ + $(\mathrm{jj})_\mathrm{VBF}$ & 6 & 230--1000 & 3\% \\
\hline
\end{tabular}
\end{center}
\end{table}

\section{Systematic uncertainties}

The main sources of systematic uncertainties arise from the assumptions in the signal model, the objects reconstruction used in the analysis, and several common experimental sources.
Theoretical uncertainties on the cross section for the heavy Higgs boson production derive from the uncertainties in the choice of the Parton Distribution Functions and $\alpha_s$, along with the renormalization and factorization scales. These are typically of the order of 6-7\% and 7–12\%, respectively, for ggF production, and 1–2\% and 2–5\%, respectively, for VBF. We also add an uncertainty on the background arising from the off-shell h(125) production, estimated using GG2VV (PHANTOM) for the ggF (VBF) case. This uncertainty is of the order of 3\% of the total background for large $m_H$ values.
The uncertainties on the lineshape of signal and interference are different depending on the production mode. For ggF, we follow the recommendation of the HXSWG~\cite{bib11}. Since there is no prescription for VF, we assign as systematic uncertainty the renormalization and factorization scale variations in PHANTOM.

A systematic uncertainty common to all decay channels is the luminosity measurement, which is 2.2\% (2.6\%) for the 7 (8) TeV data. Other uncertainties that are correlated among channels are the muon and electron reconstruction efficiencies, and the jet energy scale and resolution. Lepton fake rate is accounted for in all channels, but it is mostly relevant for the \HZZ$\to$ 2l2l' final state, where we consider leptons at lower transverse momentum with respect to the other channels.

\section{Interpretation of the results}

The statistical combination of the several final states in this analysis was developed within the {\it LHC Higgs Combination Group} by ATLAS and CMS~\cite{bib22}. To determine the limits on the model parameters as a function of $m_H$, a modified frequentist method (best known as CL$_s$) is used~\cite{bib23}. The uncertainties described in the previous section are introduced as nuisance parameters.

\subsection{Search for a SM-like heavy Higgs boson}

\begin{figure}
\includegraphics[width=0.98\textwidth]{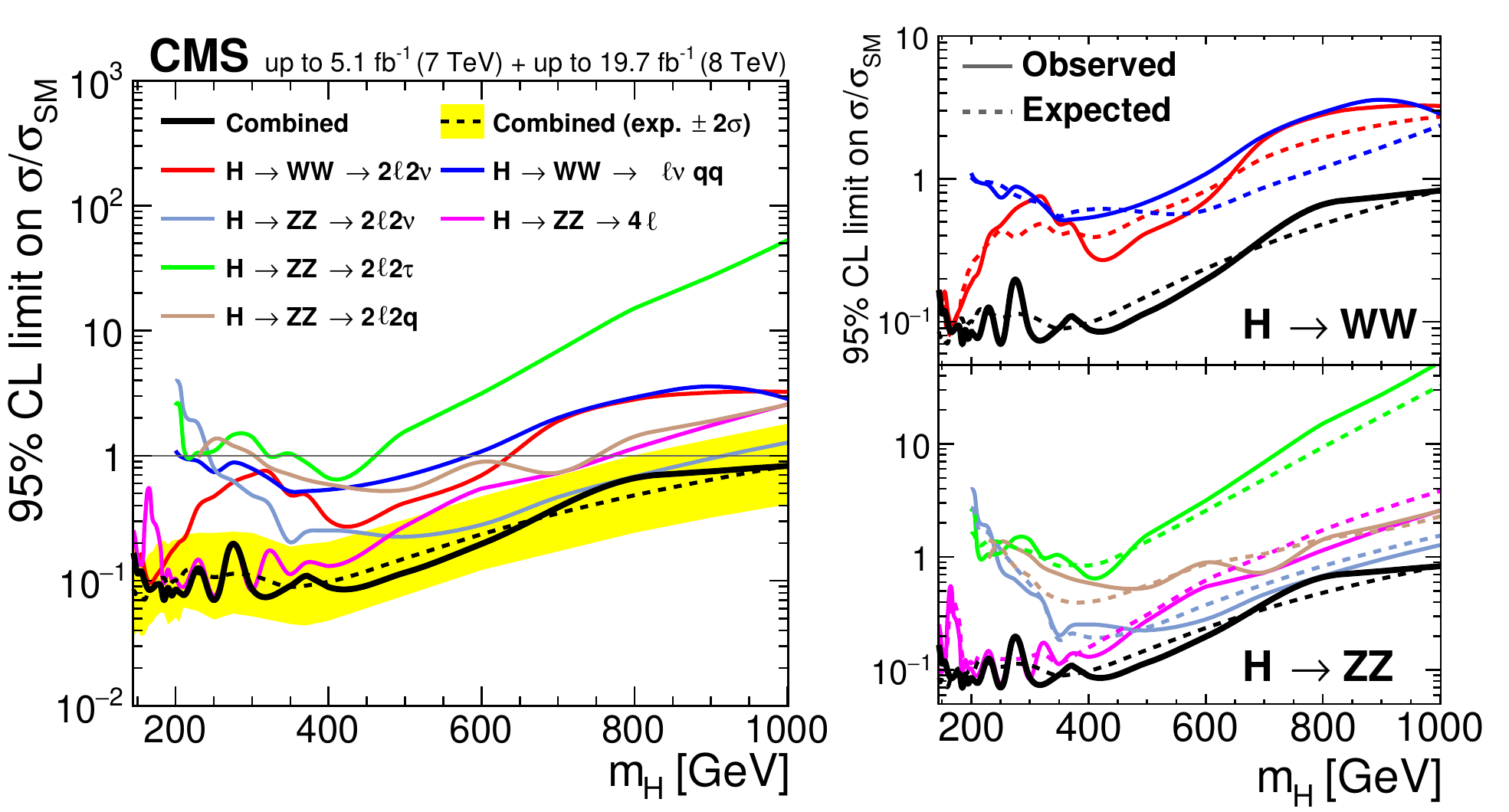}\\
\caption[]{ 95\% CL upper limits on a SM-like Higgs boson for all the final states considered and their combination. Observed and expected limits for the individual channels are shown in the right, for WW channels on top and ZZ channels on the bottom.}
\label{fig1}
\end{figure}

In Figure \ref{fig1} the combined results for the search of a heavy SM-like Higgs boson are shown. On the left, we show the observed 95\% CL limit for each final state entering in the analysis, along with the combination in black. The expected combined limit is shown as a dashed black line, along with the $\pm 2\sigma$ yellow band representing the expected interval of the limit. 
The plot on the right shows the channel by channel comparison of the expected and observed limit, using the same color legend. The top right plot refers to the WW final states, while the bottom right plot refers to the ZZ final states. While the \HZZ$\to$ 4l final state shows good sensitivity across the whole invariant mass spectrum, the \HWW$\to$ l$\nu$l$\nu$ channel is more sensitive at lower mass, while in the highest mass region \HZZ$\to$ 2l2$\nu$ is the most sensitive.

The structures present in the observed limit can be attributed to similar features in the limits of individual channels. The small excess in the combined limit around 280 GeV is present in the \HZZ$\to$ 4l and \HWW$\to$ l$\nu$l$\nu$ final states. The combined limit on the cross section times the branching ratio excludes at 95\% CL the presence of a SM-like Higgs boson across the full range of 145 $< m_H <$ 1000 GeV.

\subsection{Electroweak Singlet Interpretation}

\begin{figure}
\includegraphics[width=0.98\textwidth]{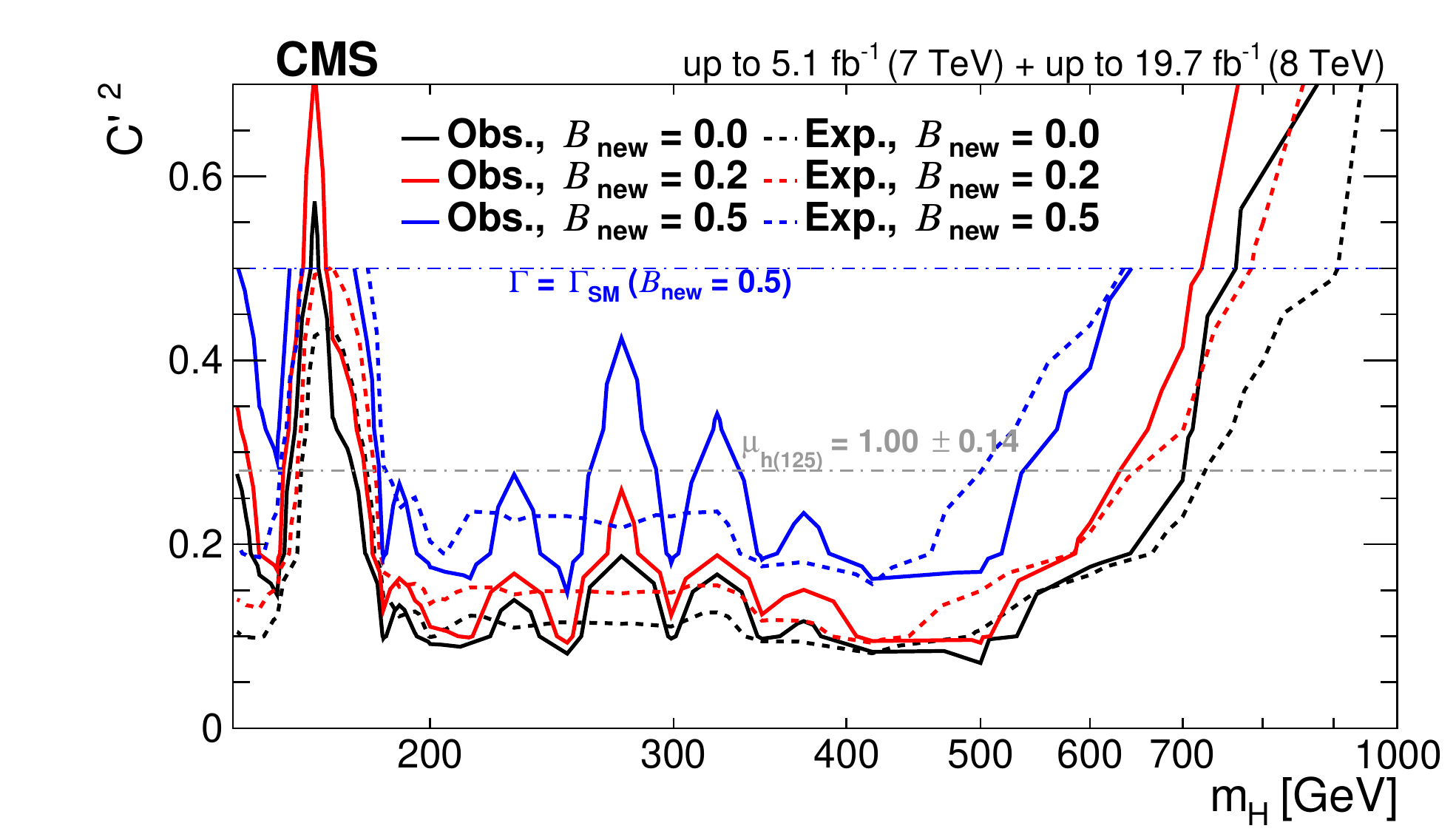}\\
\caption[]{95\% CL upper limits on C' in the EW singlet scenario, as a function of the heavy Higgs boson mass, for several $\mathcal{B}_{new}$ values. The upper blue dashed line represents where, for $\mathcal{B}_{new}$ = 0.5, the variable width of the heavy Higgs boson reaches the width of a SM-like Higgs boson. The lower grey dash-dotted line shows the indirect lower limit on C' obtained from the h(125) signal strength measurement.}
\label{fig2}
\end{figure}

In the EW singlet model, there are two parameters of interest: C', the coupling scale factor, and $\mathcal{B}_{new}$, the modifier to the total width that parametrizes additional non-SM decays for the heavy Higgs boson. In Figure~\ref{fig2} the observed and expected upper limit on C' as a function of the heavy Higgs mass are shown, for several $\mathcal{B}_{new}$ values. In the same plot, the indirect limit on C' obtained from the measurement of C for h(125) is shown. The upper blue dashed line represents where, for $\mathcal{B}_{new}$ = 0.5, the variable width of the heavy Higgs boson reaches the width of a SM-like Higgs boson.

\section{Conclusions}

We present the combination of \HZZ and \HWW decay searches for a heavy Higgs boson in the 145 $< m_H <$ 1000 GeV invariant mass range. We interpret our observed data both in a SM-like heavy Higgs scenario, and in the case of a EW singlet in addition to the 125 GeV Higgs boson. We do not observe a significant excess with respect to the expected SM background in either interpretation. In the context of a SM-like heavy Higgs boson, we are able to exclude this hypothesis in the whole mass range considered. For the EW singlet scenario, we are able to set limits on the C' parameter of the theory as a function of the heavy state mass.

\end{document}